# Hopf -bifurcation analysis of pneumococcal pneumonia with time delays

Fulgensia Kamugisha Mbabazi [1*], Joseph Y.T. Mugisha [2] and Mark Kimathi [3]

1∗ *Department of Mathematics, Pan African University Institute of Basic Sciences, Technology and Innovation, P. O. Box 62000–00200, Nairobi, Kenya*

2 *Department of Mathematics, Makerere University, P. O. Box 7062, Kampala, Uganda*

3 *Department of Mathematics, Statistics and Actuarial Sciences, Machakos University, P. O. Box, 136–90100, Machakos, Kenya*

Abstract. In this paper, a mathematical model of pneumococcal pneumonia with time delays is proposed. The stability theory of delay differential equations is used to analyze the model. The results show that the disease–free equilibrium is asymptotically stable if the control reproduction ratio $R_0$ is less than unity and unstable otherwise. The stability of equilibria with delays shows that the endemic equilibrium is locally stable without delays and stable if the delays are under conditions. The existence of Hopf-bifurcation is investigated and transversality conditions proved. The model results suggest that as the respective delays exceed some critical value past the endemic equilibrium, the system loses stability through the process of local birth or death of oscillations. Further, a decrease or an increase in the delays leads to asymptotic stability or instability of the endemic equilibrium respectively. The analytical results, are supported by numerical simulations.

## 1 Introduction

Worldwide, pneumococcal pneumonia disease continues to be a major cause of morbidity and mortality in persons of all ages and the leading cause of bacterial childhood disease, despite a century of study and the development of antibiotics and vaccination [2, 10]. Pneumococci are different, with 90 recognized serotypes; several of these serotypes are capable of causing invasive disease [4]. Pneumococcal pneumonia infections may follow a viral infection, like a cold or flu (influenza) [27], and cause the following types of illnesses depending on the affected part of the body: invasive pneumococcal diseases (IPD) such as meningitis, bacteremia and bacteremic pneumonia; lower respiratory tract infections (e.g., pneumonia), and upper respiratory tract infections (e.g., otitis media and sinusitis) [3]. The wide spread of the disease may be promoted by potentially asymptomatic persons (incubation individuals) [15, 16] and an individual remains in the exposed class for a certain latent period prior to becoming infective [13, 14].

Diseases exhibit a lot of economic burden including productivity loss, health care related expenses, losses due to disease related mortality and loss of employment [47]. Globally, an estimated 14.5 million episodes of serious pneumococcal disease occur each year among children under 5 years of age, resulting in approximately 500,000 deaths [5], most of which occur in low and middle–income countries [6, 7]. Pneumonia is the most common form of severe pneumococcal disease, accounting for 15 % of all deaths of children under 5 years and killing an estimated 922,000 in 2015, and is the leading cause of death in this age group [8].

Vaccination is a highly efficient means of preventing diseases and death [34]. A vaccine consists of a killed or weakened form or derivative of the infectious germ. Once administered to a healthy person, the vaccine activates an immune response and makes the body to assume that it is being attacked by a specific organism
[41]. Decrease of invasive pneumococcal disease (IPD) has been managed by pneumococcal conjugate vaccines (PCVs), and they are among the many ongoing stories of vaccine successes around the world. One dose of vaccine does not protect all receivers because vaccine–induced immunity is lost after some period of time [22, 23].

Time delays are significant in the transmission process of epidemics and arise due to delayed feedback especially the period for waning vaccine–induced immunity, latent period of infection, the infectious period and the immunity period [9, 12, 21]. Among the mathematical tools currently used, delay differential models with time delay have attracted attention in the field of science especially modeling infectious diseases. Delays change the dynamical systems' stability by giving rise to Hopf–bifurcations [9, 17]. Work done by researchers for example [13, 42, 43, 44, 45] demonstrate the role played by time delays in different capacities in controlling the spread of infectious diseases. Sharma et al. [1] discussed avian influenza transmission dynamics with two discrete time delays as incubation periods of avian influenza in the human and avian populations, and found out that increments in time delays occurrence results into decrease in infected human population.

In this paper, we explore the effect of two delays on pneumococcal pneumonia disease. We incorporate a time delay in the latent class because there is delayed time from the time an individual is infected and when one becomes infectious. A second time delay of seeking medical care is included in the infectious class. Not seeking medical attention leaves individuals' behaviors unchanged not to respond to existing control measures and more individuals become infected.

This paper is organized as follows. In Section 2, we present the description and formulation of the time delay model of pneumococcal pneumonia dynamics. In Section 3, we present the stability of the steady states. Existence of Hopf–bifurcation is presented in Section 4. In Section 5, numerical simulations and results of the model are presented to support the analytical findings, a discussion is given in Section 6.

## 2  Model description and formulation

We formulate a model for the dynamics of the bacterial pneumonia (pneumococcal) in a human population with the total population size at time $t$, denoted by $N(t)$. The population is sub–divided into six mutually exclusive epidemiological classes: susceptible, vaccinated, exposed, carrier and infected denoted by $S(t)$, $V(t)$, $E(t)$, $C(t)$ and $I(t)$; respectively. The mathematical formulation adopts a mass–action incidence because it's important in deciding the dynamics of epidemic models [39], where the contact rate depends on the size of the total human population [19]. We assume a continuous vaccination strategy that is received by the recruited susceptible individuals at a rate $v$, and that vaccination doesn't affect the infectious [18]. We assume vaccination is not 100% efficient, which means there is a chance of being infectious or carrier in small proportions and the force of infection for the vaccinated class is $\vartheta\beta I(t)$, where $0 \leq \vartheta < 1$ is the proportion of the sero–type not covered by vaccine [20]. The increase in the number of susceptible individuals comes from a constant recruitment $b$ through birth or migration and recovery of individuals. Several vaccines wane with time, and so vaccinated individuals return to the susceptible compartment, at a waning rate $\zeta$. The susceptible individuals become infected through a force of infection $\beta I(t)$ and move to the latent class $E(t)$.

The latent class, $E(t)$ accounts for a time delay $\tau_1 > 0$ of the exposed individuals i.e. the period between the time of an infection onset and the time of developing pneumococcal clinical symptoms (assume that an individual is infectious upon exposure to influenza A disease that promotes severe pneumococcal pneumonia). The probability (survivorship function) of an individual surviving the natural mortality through the latent period $[t - \tau_1, t]$ is $e^{-\mu\tau_1}$ and exposed individuals transfer to the infectious class at a rate *γ*. Individuals in the carrier class *C*(*t*) become symptomatic and join the infected class at a rate *ρ*.

The infectious class *I*(*t*) accounts for a time delay $\tau_2 > 0$: the time taken by infected individuals to seek medical care. We assume that infected individuals who survive the natural mortality through the infectious period [$t - \tau_2, t$] have a survivorship function $e^{-\mu\tau_2}$. Moreover, infected individuals that delay to seek medical care die of pneumococcal pneumonia at a rate *δ*. Infectious individuals upon recovery transfer to the susceptible class at a rate *φ*. All classes exhibit a per capita natural mortality rate *μ*.

The description of model variables and parameters is summarized in Table 1 and Table 2 below.

Table 1: Description of variables

| Variable | Description | [unit] |
|---|---|---|
| S(t) | Number of susceptible individuals at time *t* | individual |
| V(t) | Number of vaccinated individuals at time *t* | individual |
| E(t) | Number of asymptomatic individuals at time *t* | individual |
| C(t) | Number of individuals with one serotype not covered by the vaccine | individual |
| I(t) | Number of infectious individuals at time *t* | individual |

Table 2: Description of parameters

| Parameter | Description | value/unit | Source |
|---|---|---|---|
| b | Recruitment rate | 22 day$^{-1}$ | estd |
| v | Effective vaccination rate | $2.53 \times 10^{-5}$ | [28] |
| γ | Transfer rate from *E* to *I* class | $3.3333 \times 10^{-1}$ day$^{-1}$ | assumed |
| μ | Natural mortality rate from causes unrelated to the infection | $2.0547 \times 10^{-3}$ | [29] |
| δ | Disease–induced mortality rate | $3.3 \times 10^{-1}$ day$^{-1}$ | [33] |
| ρ | Progression rate from *C* to *I* class | $1.096 \times 10^{-2}$ day$^{-1}$ | [30] |
| φ | Per capita rate of recovery | $3.5714 \times 10^{-2}$ day$^{-1}$ | [11] |
| ζ | Waning rate of vaccine | $5.4794 \times 10^{-4}$ day$^{-1}$ | [11] |
| ϑ | Proportion of the sero–type not covered by vaccine | 0.54 | [46] |
| β | Transmission coefficient | $1.0102 \times 10^{-4}$ day$^{-1}$ | assumed |
| $\tau_1$ | Delay for the incubating individuals | 1–3 days | [31] |
| $\tau_2$ | Delay in seeking medical care | 2 days | [32] |

The compartmental diagram of the model is shown in Figure.1.

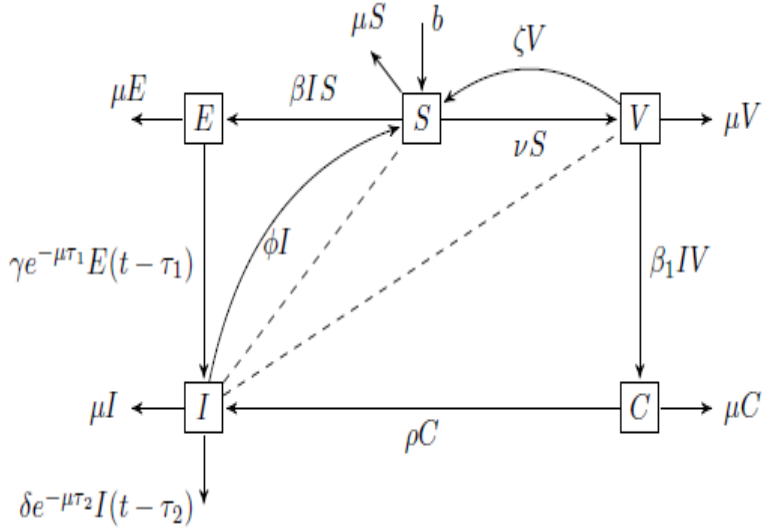

Figure 1: A schematic diagram showing the dynamics of pneumococcal pneumonia. The dotted lines represent contacts made by individuals in the respective classes and the solid lines show transfer from one class to another. Based on the description of model variables, parameters and assumptions above, the dynamics of the model are governed by the following differential equations:

$$\dot{S}(t) = b + \zeta V(t) + \phi I(t) - (v + \mu + \beta I(t))S(t),$$

$$\dot{V}(t) = vS(t) - (\mu + \zeta)V(t) - \beta_1 I(t)V(t),$$

$$\dot{E}(t) = \beta I(t)S(t) - \gamma e^{-\mu \tau_1} E(t - \tau_1) - \mu E(t), \qquad (1)$$

$$\dot{C}(t) = \beta_1 I(t)V(t) - (\rho + \mu)C(t),$$

$$\dot{I}(t) = \rho C(t) + \gamma e^{-\mu \tau_1} E(t - \tau_1) - \delta e^{-\mu \tau_2} I(t - \tau_2) - (\mu + \phi)I(t).$$

where, $\beta_1 = \vartheta \beta$.

## 2.1 Positivity of solutions

System (1) is a representation of the dynamics of the human populations, thus it is required that all solutions are non–negative. We use the approach of Bodna [35] and Yang et al. [36], we let $C$ be a Banach space of continuous real valued functions $\psi:[-\tau,0]\to\Re_+^5$ equipped with the supremum norm, $\|\psi\|_C=\sup_{t\in[-\tau,0]}\{|\psi_1|,|\psi_2|,|\psi_3|,|\psi_4|,|\psi_5|\}$. The initial conditions of system (1) are represented by

$$S(t)=\psi_1(t),\ V(t)=\psi_2(t),\ E(t)=\psi_3(t),\ C(t)=\psi_4(t),\ I(t)=\psi_5(t),\ -\tau\leq t\leq 0, \qquad (2)$$

where $\tau=\max\{\tau_1,\tau_2\}$ and $\psi=(\psi_1,\psi_2,\psi_3,\psi_4,\psi_5)^T\in C$, such that $\psi_i(t)=\psi_i(0)\geq 0\,(i=1,2,3,4,5)$. The following Lemma establishes the positivity of the solutions of system (1).

**Lemma 2.1.** *Any solution of trajectories* (1) *with* $\psi_i(t)>0; t\in[-\tau,0]$ *remains positive whenever it exists*.

**Proof.** Suppose $S(t)$ was to lose positivity on some local existence interval $[0, T)$ for some constant $T > 0$, there would be a time at $t_1 = \sup\{t > 0: S(t) > 0\}$ such that $S(t_1) = 0$. From the first equation of system (1), it follows that

$$b+\zeta V(t)+\phi I(t)-(\nu+\mu)S(t)-\beta I(t)S(t)>0.$$

This implies that $S(t) < 0$ for $t \in (t_1 - \varepsilon, t_1)$, where $\varepsilon$ is an arbitrary small positive constant. This leads to a contradiction, it thus follows that $S(t)$ is always positive. Hence from the fundamental theory of differential equations, it is shown that there exists a unique solution for $S(t)$ of system (1) with initial data in $\mathbb{R}_+^5$ as follows

$$\frac{d}{dt}\left(S(t)e^{\int_0^t(\nu+\mu+\beta I(t))d\xi}\right)=e^{\int_0^t(\nu+\mu+\beta I(t))d\xi}(b+\zeta V(t)+\phi I(t)),$$

$$S(t)=\int_0^t\left((b+\zeta V(\sigma)+\phi I(\sigma))e^{-(\nu+\mu)t-\int_\sigma^t\beta I(\xi)d\xi}\right)d\sigma+\psi_1(0)e^{-(\nu+\mu)t-\int_0^t\beta I(\xi)d\xi},$$

Therefore, $S(t_1)=\psi_1(0)e^{-(\nu+\mu)t_1-\int_0^{t_1}\beta I(\xi)d\xi}+\int_0^{t_1}\left((b+\zeta V(\sigma)+\phi I(\sigma))e^{-(\nu+\mu)t_1-\int_\sigma^{t_1}\beta I(\xi)d\xi}\right)d\sigma>0.$ (3)

Since $S(t_1) > 0$, then $S(t) > 0$, $t \geq 0$. This completes the proof.

Similarly, it can be shown that

$$V(t_2) = \psi_2(0)e^{-(\mu+\zeta)t_2 - \int_0^{t_2}\beta I(\xi)d\xi} + \int_0^{t_2} e^{\int_\sigma^{t_2}(\mu+\zeta+\beta_1 I(\xi))d\xi} vS(\sigma)d\sigma > 0. \tag{4}$$

$$E(t_3) = e^{-\mu t_3}\psi_3(0) + e^{-\mu t_3}\left(\int_0^{t_3} e^{\mu\xi}\left(\beta I(\xi)S(\xi) - \gamma e^{-\mu\tau_1}E(\xi-\tau_1)\right)d\xi\right) > 0. \tag{5}$$

$$C(t_4) = e^{-(\rho+\mu)t_4}\left(\psi_4(0) + \int_0^{t_4}\left(\beta_1 I(\xi)V(\xi)\right)e^{(\rho+\mu)\xi}d\xi\right) > 0. \tag{6}$$

and

$$I(t_5) = \psi_5(0)e^{-(\mu+\phi)t_5} + e^{-(\mu+\phi)t_5}\left(\int_0^{t_5}\left(\rho C(\sigma) + \gamma e^{-\mu\tau_1}E(\sigma-\tau_1) - \delta e^{-\mu\tau_2}I(\sigma-\tau_2)\right)e^{(\mu+\phi)\sigma}\right)d\sigma. \tag{7}$$

Therefore, from the above integral forms of equations (3) to (7) all solution trajectories are positive for all time $t > 0$ on $[0, +\infty]$.

## 2.2 Boundedness

For boundedness of system (1) with initial condition (2), we consider the following Lemma:

**Lemma 2.2.** *The closed set*

$$\Omega_d = \{S(t), V(t), E(t), C(t), I(t), R(t)\} \in \mathfrak{R}_+^5 : 0 \leq S(t), V(t), E(t), C(t), I(t);$$

$$S(t) + V(t) + E(t) + C(t) + I(t) \leq \frac{b}{\mu}$$

*is positively invariant and absorbing with respect to the set of DDE's* (1).

**Proof.** Summing all equations in system (1), yields:

$$\frac{dN}{dt} = b - \mu N(t) - \delta e^{-\mu\tau_2}I(t).$$

Therefore, $\frac{dN}{dt} \leq b - \mu N(t)$ which implies that $\frac{dN}{dt} \leq 0$ if $N(t) \geq \frac{b}{\mu}$. Using the standard comparison test in [25], we get $N(t) \leq N(0)e^{-\mu t} + \frac{b}{\mu}(1 - e^{-\mu t})$. Particularly, $N(t) \leq \frac{b}{\mu}$ if $N(0) \leq \frac{b}{\mu}$ for all time $t > 0$, hence $\Omega_d$ is positively invariant. Further, if $N(t) \geq \frac{b}{\mu}$, then either the solution enter at finite time or $N(t)$ is close to $\frac{\pi}{\mu}$ and the infected variables $E$, $C$ and $I$ tend to zero. Therefore, $\Omega_d$ is attracting implying that all solutions in $\Re_+^5$ finally enter $\Omega_d$ consequently, in $\Omega_d$, system (1) is mathematically and epidemiologically well–posed.

## 2.3 The control reproduction ratio

The basic reproduction ratio identifies the number of secondary infections from the infected source and plays an important role in understanding the development of epidemics with a vaccination program in place. The control reproduction ratio $R_0$ is computed using an approach in [24] and is given by $R_0 = R_0^u + R_0^v$.

where,

$$R_0^u = \frac{\beta \gamma e^{-\mu \tau_1} S^0}{(\mu + \gamma e^{-\mu \tau_1})(\phi + \mu + \delta e^{-\mu \tau_2})}, \quad R_0^v = \frac{\beta \rho \vartheta V^0}{(\rho + \mu)(\phi + \mu + \delta e^{-\mu \tau_2})},$$ provided the validity of

$(\mu + \zeta)(\nu + \mu) > \zeta \nu$ holds. The quantity $R_0^u$ measures the expected number of secondary cases generated by an index case for the susceptible individuals and $R_0^v$ represents new cases arising from the vaccination program.

**Remark 2.1.** The control reproduction ratio with no delays ($\tau_1 = 0$, $\tau_2 = 0$) are given by

$$R_0^u = \frac{\beta \gamma S^0}{(\mu + \gamma)(\phi + \mu + \delta)} \text{ and } R_0^v = \frac{\beta \rho \vartheta V^0}{(\rho + \mu)(\phi + \mu + \delta)}.$$

## 3 Stability of equilibria

Let $(S^*, V^*, E^*, C^*, I^*)$ be the corresponding partial populations at the eventual equilibrium point. Given that the values of the partial populations at the equilibrium are stable, the delay–dependency vanishes so that

$$\lim_{t \to \infty} I(t - \tau_2) = \lim_{t \to \infty} I(t) = I^* \text{ and } \lim_{t \to \infty} E(t - \tau_1) = \lim_{t \to \infty} E(t) = E^*,$$ such that at equilibrium, we have

$$b + \zeta V^* + \phi I^* - (\nu + \mu + \beta I^*)S^* = 0,$$

$$\nu S^* - (\mu + \zeta)V^* - \beta_1 I^* V^* = 0,$$

$$\beta I^* S^* - (\gamma e^{-\mu \tau_1} + \mu)E^* = 0, \qquad (8)$$

$$\beta_1 I^* V^* - (\rho + \mu)C^* = 0,$$

$$\rho C^* + \gamma e^{-\mu \tau_1} E^* - (\mu + \delta e^{-\mu \tau_2} + \phi)I^* = 0,$$

$$\dot{S}^* + \dot{V}^* + \dot{E}^* + \dot{C}^* + \dot{I}^* = b - \mu N^* - \delta e^{-\mu \tau_2} I^* = 0.$$

Hence, from system (8), we obtain the disease-free equilibrium $P_0 = (S^0, V^0, 0, 0, 0)$,

$$\text{where } S^0 = \frac{b(\mu + \zeta)}{(\mu + \zeta)(\nu + \mu) - \zeta \nu}, V^0 = \frac{b\nu}{(\mu + \zeta)(\nu + \mu) - \zeta \nu}, \qquad (9)$$

provided $(\mu + \zeta)(\nu + \mu) > \zeta \nu.$

It should be noted that for $\nu > 0$, the disease–free equilibrium is biologically feasible for any epidemiological parameters, whereas in the absence of vaccination strategy, i.e. for $\nu = 0$, $E_0$ is only feasible for epidemiological parameters in the susceptible class. From system (8) the endemic equilibrium $P^* = (S^*, V^*, E^*, C^*, I^*)$ is given as

$$S^* = \frac{b+\zeta V^* + \phi I^*}{\nu+\mu+\beta I^*}, \quad V^* = \frac{\nu(b+\phi I^*)}{(\nu+\mu+\beta I^*)(\mu+\zeta+\beta_1 I^*)-\nu\zeta}, \quad E^* = \frac{\beta(\zeta\nu+a_1)(bI^* + \phi I^{*2})}{a_1(\gamma e^{-\mu\tau_1}+\mu)(\nu+\mu+\beta I^*)},$$

$$C^* = \frac{\nu\beta_1 I^*(b+\phi I^*)}{a_1(\rho+\mu)}, \quad I^* = I^*. \tag{10}$$

where $a_1 = (\nu+\mu+\beta I^*)(\mu+\zeta+\beta_1 I^*)-\nu\zeta$.

### 3.1 Local stability of the disease–free equilibrium point

Suppose that $P_0 = (S^0, V^0, 0, 0, 0)$ is a disease–free equilibrium point of system (1), then the linearization matrix $J_{P_0}$ is given by

$$J_{P_0} = \begin{vmatrix} -(\mu+\nu) & \zeta & 0 & 0 & \phi-\beta S^0 \\ \nu & -(mu+\zeta) & 0 & 0 & -\beta\vartheta V^0 \\ 0 & 0 & -\mu & 0 & \beta S^0 \\ 0 & 0 & 0 & -(\rho+\mu) & \beta\vartheta V^0 \\ 0 & 0 & 0 & \rho & -(\mu+\phi) \end{vmatrix} = 0.$$

Clearly $y_1 = -\mu$ is one of the negative roots (eigenvalues) that guarantee local stability of the disease–free equilibrium $P_0$. The remaining eigenvalues are obtained from the characteristic polynomial given by

$$g(y) = y^4 + e_3 y^3 + e_2 y^2 + e_1 y + e_0 = 0, \tag{11}$$

where

$e_3 = 4\mu+\nu+\zeta+\rho+\phi, e_2 = (\mu+\zeta)(2\mu+\rho+\nu)+(\mu+\phi)(2\mu+\nu+\zeta)-\beta\vartheta\rho V^0,$

$e_1 = (\mu+\zeta)(\mu+\rho)(2\mu+\nu+\phi)+(\mu+\nu)(\mu+\phi)(2\mu+\zeta+\rho)-\zeta\nu(2\mu+\rho+\phi)-\beta\rho\vartheta(2\mu+\nu+\zeta),$

$e_0 = (\mu+\nu)(\mu+\zeta)\big((\rho+\mu)(\mu+\phi)-\beta\rho\vartheta V^0\big)+\zeta\nu(\beta\rho\vartheta V^0-(\rho+\mu)(\mu+\phi)).$

Thus, computing the roots of polynomial (11) gives

$$y_2 = -\mu,\ y_3 = -(\mu+\zeta+\nu),\ y_4 = -\frac{1}{2}\left((2\mu+\rho+\phi)+\sqrt{(\rho-\phi)^2+4\beta\rho\vartheta V^0}\right),$$

$$y_5 = -\frac{1}{2}\left((2\mu+\rho+\phi)-\sqrt{(\rho-\phi)^2+4\beta\rho\vartheta V^0}\right).$$

Since the rest of the roots are negative, root $y_5$ is negative provided $(\mu+\phi)(\rho+\mu) > \beta\rho\vartheta V^0$, holds implying that

$$R_0^v = \frac{\beta\rho\vartheta b\nu}{(\mu+\phi)(\rho+\mu)\left((\mu+\zeta)(\nu+\mu)-\zeta\nu\right)} < 1.$$

Thus we have the result below

**Proposition 3.1.** *The disease–free equilibrium $P_0$ is locally asymptotically stable if the control reproduction ratio $R_0 < 1$, whenever conditions $(\mu+\zeta)(\mu+\nu) > \zeta\nu$ and $R_0^v < 1$ are satisfied, and unstable otherwise.*

To illustrate the stability of disease–free equilibrium, we use parameter values in Table 2 with corresponding population estimates of $S(0) = 10604$, $V(0) = 103$, $E(0) = C(0) = I(0) = 0$ and the resulting simulation is given in Figure 2.

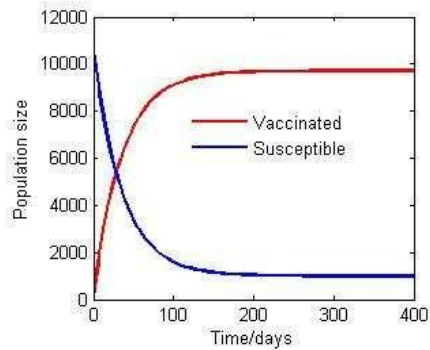

Figure 2: Simulation of model (1), the disease–free equilibrium, with populations parameters: $\varphi = 3.57144 \times 10^{-1}$, $\beta = 1.0102 \times 10^{-5}$, $\gamma = 3.3333 \times 10^{-2}$ (with $R_0 = 0.7873, R_0^u = 0.1382, R_0^v = 0.6490$).

The biological implication of Proposition 3.1, means that in the long run the vaccinated and susceptible populations will be stable and pneumococcal pneumonia will be under control.

## 3.2 The transcendental equation

We obtain the expression for the transcendental equation by linearizing system (1) around

$P^* = (S^*, V^*, E^*, C^*, I^*)$, to obtain

$$\begin{pmatrix} \dot{S}(t) \\ \dot{V}(t) \\ \dot{E}(t) \\ \dot{C}(t) \\ \dot{I}(t) \end{pmatrix} = \begin{pmatrix} a_1 & a_2 & 0 & 0 & a_3 \\ a_4 & a_5 & 0 & 0 & a_6 \\ a_7 & 0 & a_8 & 0 & a_9 \\ 0 & a_{10} & 0 & a_{11} & a_{12} \\ 0 & 0 & 0 & a_{13} & a_{14} \end{pmatrix} \begin{pmatrix} S(t) \\ V(t) \\ E(t) \\ C(t) \\ I(t) \end{pmatrix} + \begin{pmatrix} 0 & 0 & 0 & 0 & 0 \\ 0 & 0 & 0 & 0 & 0 \\ 0 & 0 & a_{15} & 0 & 0 \\ 0 & 0 & 0 & 0 & 0 \\ 0 & 0 & a_{16} & 0 & a_{17} \end{pmatrix} \begin{pmatrix} 0 \\ 0 \\ E_{\tau_1} \\ 0 \\ I_{\tau_2} \end{pmatrix}, \qquad (12)$$

$a_1 = -((\mu+v)+\beta I^*), a_2 = \zeta, a_3 = \phi - \beta S^*, a_4 = v, a_5 = -((\mu+\zeta)+\beta \vartheta I^*), a_6 = -\beta \vartheta V^*,$

$a_7 = \beta I^*, a_8 = -\mu, a_9 = \beta S^*, a_{10} = \beta \vartheta I^*, a_{11} = -(\rho+\mu), a_{12} = \beta \vartheta V^*, a_{13} = \rho,$

$a_{14} = -(\mu+\phi), a_{15} = -\gamma e^{-\mu\tau_1}, a_{16} = \gamma e^{-\mu\tau_1}, a_{17} = -\delta e^{-\mu\tau_2}, E_{\tau_1} = E(t-\tau_1)$ and $I_{\tau_2} = I(t-\tau_2)$.

The variational matrix of (12) is given by

$$g(\lambda, e^{-\lambda\tau_1}, e^{-\lambda\tau_2}) = \begin{vmatrix} a_1 - \lambda & a_2 & 0 & 0 & a_3 \\ a_4 & a_5 - \lambda & 0 & 0 & a_6 \\ a_7 & 0 & a_8 - \gamma e^{-(\mu+\lambda)\tau_1} - \lambda & 0 & a_9 \\ 0 & 0 & 0 & a_{11} - \lambda & a_{12} \\ 0 & 0 & \gamma e^{-(\mu+\lambda)\tau_1} & a_{13} & a_{14} - \delta e^{-(\mu+\lambda)\tau_2} - \lambda \end{vmatrix} = 0,$$

Then, we obtain the transcendental equation of the linearized system at $P^*$:

$$g(\lambda, e^{-\lambda\tau_1}, e^{-\lambda\tau_2}) = \lambda^5 + k_4\lambda^4 + k_3\lambda^3 + k_2\lambda^2 + k_1\lambda + k_0 + (\lambda^4 + l_3\lambda^3 + l_2\lambda^2 + l_1\lambda + l_0)\gamma e^{-(\mu+\lambda)\tau_1}$$

$$(\lambda^4 + m_3\lambda^3 + m_2\lambda^2 + m_1\lambda + m_0)\delta e^{-(\mu+\lambda)\tau_2} + (\lambda^3 + n_2\lambda^2 + n_1\lambda + n_0)\gamma\delta e^{-(\mu+\lambda)(\tau_1+\tau_2)} = 0. \quad (13)$$

with coefficients of the transcendental equation (13) given in Appendix A.1.

## 3.3 Delay–free system

Here, to show the local stability of $P^*$, we consider a situation where there are no delays during the latent period ($\tau_1 = 0$) and in seeking medical care ($\tau_2 = 0$). By letting $\tau_1 = \tau_2 = 0$, equation (13) reduces to:

$$g(\lambda) = \lambda^5 + b_4\lambda^4 + b_3\lambda^3 + b_2\lambda^2 + b_1\lambda + b_0 = 0. \tag{14}$$

with coefficients of polynomial equation in Appendix A.2.

**Proposition 3.2**. *The endemic equilibrium $P^*$ is locally asymptotically stable in the absence of delays $\tau_1 = \tau_2 = 0$, iff the following Routh–Hurwitz conditions are satisfied:*

$b_0 > 0, b_4 b_3 - b_2 > 0, b_2(b_4 b_3 - b_2) - b_4(b_4 b_1 - b_0) > 0$ and

$b_1(b_2(b_4 b_3 - b_2) - b_4(b_4 b_1 - b_0)) - b_0(b_3(b_4 b_3 - b_2) - (b_4 b_1 - b_0)) > 0$, with $b_4, b_3, b_2, b_1$ and $b_0$ defined in Appendix A.2.

Numerically, using parameter values in Table 2 the characteristic equation (14) is given as $\lambda^5 + 0.7364\lambda^4 - 148.4\lambda^3 - 4.9408\lambda^2 - 0.3965\lambda - 0.0001806 = 0$.

The resulting eigenvalues are given by:

$\lambda_1 = 11.8, \lambda_2 = -0.0005, \lambda_3 = -12.5357, \lambda_{4,5} = -0.01641 \pm 0.4885i.$

Since there exists a positive root for model (1), there is a stability change from unstable to stable of the endemic equilibrium point $P^* = (S^*, V^*, E^*, C^*, I^*) = (2099, 6, 54, 2, 100)$ that gives rise to a Hopf–bifurcation.

## 4 Existence of Hopf–bifurcation

Under this subsection, we discuss the stability of the endemic equilibrium point of model (1). We use the approach of Song and Wei [38] to prove the conditions for continuation of unstable or stable switches at the endemic equilibrium point. By choosing time delay $\tau = \max = \{\tau_1, \tau_2\}$ as the bifurcation parameter.

## 4.1 Delay only in latent period ($\tau_1 > 0, \tau_2 = 0$)

In such a situation the characteristic equation (13) reduces to

$$\lambda^5 + k_4\lambda^4 + k_3\lambda^3 + k_2\lambda^2 + k_1\lambda + k_0 + (\gamma\lambda^4 + h_3\lambda^3 + h_2\lambda^2 + h_1\lambda + h_0)e^{-(\mu+\lambda)\tau_1} = 0, \qquad (15)$$

where

$$q = e^{-\mu\tau_1}, h_0 = q\gamma(l_0 + n_0\delta), h_1 = q\gamma(l_1 + n_1\delta), h_2 = q\gamma(l_2 + n_2\delta), h_3 = q\gamma(l_3 + \delta).$$

Suppose the endemic equilibrium of system (1) is stable in the absence of delay ($\tau_2$) to seek medical care, implying that $Re(\lambda) = \xi(0) < 0$. The bifurcation value of $\tau_{1_0} > 0$ occurs when $\lambda(\tau_{1_0}) = \xi(\tau_{1_0}) + iw(\tau_{1_0})$ is purely imaginary, for $\xi(\tau_{1_0}) = 0$. Hence, defining the eigenvalue $\lambda = wi$, with infection rate oscillation frequency ($w > 0$) and making a substitution in (15) and expressing the exponential in terms of trigonometric ratios, we get,

$$\begin{aligned} Im &:= a_1 \cos w\tau_1 + a_2 \sin w\tau_1 = R_1, \\ Re &: a_2 \cos w\tau_1 - a_1 \sin w\tau_1 = R_2, \end{aligned} \qquad (16)$$

Where

$$a_1 = w(h_1 - h_3 w^2), a_2 = w(\gamma w^3 - h_2 w) + h_0, R_1 = w(k_3 w^2 - (w^4 + k_1)), R_2 = k_2 w^2 - (k_4 w^4 + k_0).$$

By eliminating $\tau_1$ from equation (15), squaring and adding these two equations and putting $w^2 = z$, we obtain the Hopf frequency below:

$$H(z) = z^5 + B_4 z^4 + B_3 z^3 + B_2 z^2 + B_1 z + B_0 = 0, \qquad (17)$$

where

$$B_4 = k_4 - 2(k_3 + \gamma), B_3 = k_3^2 + 2[(k_1 + 2h_2\gamma) - (k_2 k_4 + h_3^2)], B_2 = k_2 + 2[(2h_1 h_3 + k_4 k_0) - (k_1 k_3 + 2h_0\gamma + h_2^2)],$$
$$B_1 = k_1^2 + 2[2h_0 h_2 - (h_1^2 + k_2 k_0)], B_0 = 2h_0^2 + k_0^2.$$

The two Propositions about stability and critical delay in Wesley et al. [37] are written as lemmas

**Lemma 4.1.** *If the $B_i (i=0,1,2,3,4)$ guarantee the Routh–Hurwitz criteria, then all eigenvalues of (17) have negative real parts for all delay $\tau_1 \geq 0$. Thus the endemic equilibrium $P^*$ if it exists is locally asymptotically stable whenever $\tau_1 \geq 0$, provided the endemic steady state is stable in the absence of the latent period delay, specifically $\tau_1$ won't affect the stability of the dynamical system, for equation (17) without positive real roots.*

**Lemma 4.2.** *If $B_i (i=0,1,2,3,4)$ do not satisfy Routh–Hurwitz criteria, thus $B_0 < 0$ or $B_0 > 0$ implies that (24) has at least one positive root and suppose that, it has a pair of imaginary roots say $\pm i w_{1_0}$ for such a value of $w_{1_0}$*

Consequently, to obtain the main results in this paper, we assume equation (17) has at least one positive root $w_{10}$. By squaring and summing together the imaginary and real parts in equation (16), we get

$$\tau_1 = \frac{1}{w} \arccos\left( \frac{w^2(h_1 - h_3 w^2)(k_3 w^2 - (w^4 + k_1)) + (w(\gamma w^3 - h_2 w) + h_0)(k_2 w^2 - (k_4 w^4 + k_0))}{w(h_1 - h_3 w^2)^2 + (w(\gamma w^3 - h_2 w) + h_0)^2} \right) + \frac{2n\pi}{w}. \tag{18}$$

By denoting

$$\tau_{1_n}^{(m)} = \frac{1}{w_{1_n}} \arccos\left( \frac{w_{1_n}^2(h_1 - h_3 w_{1_n}^2)(k_3 w_{1_n}^2 - (w_{1_n}^4 + k_1)) + (w_{1_n}(\gamma w_{1_n}^3 - h_2 w_{1_n}) + h_0)(k_2 w_{1_n}^2 - (k_4 w_{1_n}^4 + k_0))}{w_{1_n}(h_1 - h_3 w_{1_n}^2)^2 + (w_{1_n}(\gamma w_{1_n}^3 - h_2 w_{1_n}) + h_0)^2} \right) + \frac{2n\pi}{w_{1_n}},$$

$m = 1,2,\ldots,\tilde{m}, n \in \square$

This allows us to define

$$\tau_{1_0} = \tau_{n_0}^{(0)} = \min_{1 \leq n \leq 5}\{\tau_{1_n}^{(0)}\}, w_{1_0} = w_{n_0}. \tag{19}$$

and state the result as follows:

**Lemma 4.3.** *If $\tau_{1_0}$ and $w_{1_0}$ are defined as (19) and $H'(z = w^2) > 0$. The endemic equilibrium point $P^*$ is linearly asymptotically stable for $\tau_1 < \tau_{1_0}$, unstable for $\tau_1 > \tau_{1_0}$ and undergoes a Hopf–bifurcation at $\tau_1 = \tau_{1_0}$.*

To ensure the occurrence of the Hopf bifurcation, it is desirable to verify the transversality condition. Without loss of generality, the delay $\tau_1$ is chosen as the bifurcation parameter. The essential condition for existence of the Hopf–bifurcation is that the threshold eigenvalues traverse the imaginary axis with non–zero velocity.

**Proposition 4.1.** *If* $\Phi_2(w_{l_0}) > 0$, *where* $\Phi_2(w_{l_0})$ *satisfies* (24), *system* (1) *under goes a Hopf–bifurcation at the endemic equilibrium as* $\tau_1$ *increases through* $\tau_1$.

**Proof** (Transversality condition for Hopf–bifurcation)

Differentiating equation (15) with respect to $\tau_1$ we obtain

$$\frac{d\tau_1}{d\lambda} = \frac{(5\lambda^4 + 4k_4\lambda^3 + 3k_3\lambda^2 + k_1)e^{\lambda\tau_1} + (4\gamma\lambda^3 + 3h_3\lambda^2 + 2h_2\lambda + h_1)}{\gamma\lambda^5 + h_3\lambda^4 + h_2\lambda^3 + h_1\lambda^2 + h_0\lambda} - \frac{\tau_1}{\lambda}, \quad (20)$$

$$\text{sign}\left[\frac{(d\text{Re}\lambda)}{d\tau_1}\right]_{\tau_1=\tau_{l_0}} = \text{sign}\left[\text{Re}(\frac{d\lambda}{d\tau_1})^{-1}\right]_{\lambda=iw_{l_0}} = \text{sign}[\text{Re}N_1] + \text{sign}[\text{Re}N_2]$$

$$= \text{sign}\left[\frac{c_3(c_1\cos w\tau_1 + c_2\sin w\tau_1) + c_4(c_1\sin w\tau_1 + c_2\cos w\tau_1) + (h_1 - 3h_3w^2) + c_4w(2h_2 - 4\gamma w^3)}{c_3^2 + c_4^2}\right]. \quad (21)$$

with

$$N_1 = \frac{c_3(c_1\cos w\tau_{l_0} + c_2\sin w\tau_{l_0}) + c_4(c_1\sin w\tau_{l_0} + c_2\cos w\tau_{l_0})}{c_3^2 + c_4^2}$$

$$+ \frac{i(c_3(c_2\cos w\tau_1 + c_{l_0}\sin w\tau_{l_0}) - c_4(c_2\sin w\tau_{l_0} + c_1\cos w\tau_{l_0})}{c_3^2 + c_4^2},$$

$$N_2 = \frac{(h_1 - 3h_3w^2) + c_4w(2h_2 - 4\gamma w^3) + i(c_3w(2h_2 - 4\gamma w^2) + (h_1 - 3h_3w^2))}{c_3^2 + c_4^2}.$$

**Remark 4.1.** For any linear combination of a sine and cosine of equal periods is equal to a single sine with the same period however, with an infection rate oscillation phase shift $\Psi$ [40]. Therefore, we get

$$\text{sign}\left[\frac{(d\text{Re}\lambda)}{d\tau_1}\right]_{\tau_1=\tau_{l_0}} = \text{sign}\left[\text{Re}(\frac{d\lambda}{d\tau_1})^{-1}\right]_{\lambda=iw_{l_0}} = \frac{D_0\sin(w\tau_1 + \Psi_2) + (h_1 - 3h_3w^2) + c_4w(2h_2 - 4\gamma w^3)}{c_3^2 + c_4^2},$$

where

$$c_1 = 5w^4 - 3k_3w^2 + k_1, c_2 = 4k_4w^3, c_3 = w(\gamma w^4 - h_2w^2 + h_0), c_4 = w^2(h_3w^2 - h_1),$$

$$D_0 = \sqrt{(c_3c_1 + c_2c_1)^2 + (c_3c_2 + c_4c_1)^2}, D_0^1 = c_4w_{1_0}(2h_2 - 4\gamma w_{1_0}^3), \Psi_2 = \arctan\frac{c_3c_1 + c_4c_2}{c_3c_2 + c_4c_1}$$

Let $\Phi_2(w_{1_0}) = D_0 \sin(w_{1_0}\tau_{1_0} + \Psi) + (h_1 - 3h_3w_{1_0}^2) + c_4w_{1_0}(2h_2 - 4\gamma w_{1_0}^3) > 0,$

if conditions $(w_{1_0}\tau_{1_0} + \Psi_2) \in (\pi, \frac{\pi}{2}), h_1 > 3h_3w_{1_0}^2$ and $h_2 > 2\gamma w_{1_0}^2$ hold. Clearly

$$\text{sign}\left[\frac{(\text{dRe}\lambda)}{d\tau_1}\right]_{\tau_1=\tau_{1_0}} = \text{sign}\left[\text{Re}(\frac{d\lambda}{d\tau_1})^{-1}\right]_{\lambda=iw_{1_0}} = \frac{D_0\sin(w_{10}\tau_{1_0}+\Psi_2)+(h_1-3h_3w_{1_0}^2)+D_0^1}{c_3^2+c_4^2},$$

has the same sign as $\Phi_2(w_{1_0})$. This completes the proof.

Therefore, Proposition 4.1 implies that given $m > 0$, the eigenvalue $\lambda_m(\tau_1)$ of the characteristic equation (15) close to $\tau_{1_m}$ crosses the imaginary axis from the left to the right as $\tau_1$ continuously changes from a value less than $\tau_{1_m}$ to one greater than $\tau_{1_m}$.

## 4.2 Delay only in seeking medical care by the infectious ($\tau_1 = 0, \tau_2 > 0$)

To understand the influence of time delay in seeking medical care, we set $\tau_1 = 0$ in equation (13) yielding:

$$g(\lambda, e^{-\lambda\tau_2}) = \lambda^5 + p_4\lambda^4 + p_3\lambda^3 + p_2\lambda^2 + p_1\lambda + p_0$$

$$+(q_4\lambda^4 + q_3\lambda^3 + q_2\lambda^2 + q_1\lambda + q_0)p\gamma\delta e^{-\lambda\tau_2} = 0, \qquad (22)$$

where,

$$p = e^{-\mu\tau_2}, p_4 = k_4 + \gamma, p_3 = k_3 + l_3\gamma, p_2 = k_2 + \gamma, p_1 = k_1 + l_1\gamma, p_0 = k_0 + l_0\gamma,$$
$$q_4 = \delta p, q_3 = (m_3 + \gamma + \delta)p, q_2 = (m_2\delta + n_2\gamma\delta), q_1 = (m_1 + n_1\gamma\delta)p, q_0 = (m_0 + n_0\gamma\delta)p$$

**Proposition 4.2.** *The endemic equilibrium point $P^*$ is locally asymptotically stable (LAS) for $\tau_2 < \tau_{2_0}$ where, $\tau_{2_0}$ is the minimum positive value of*

$$\bar{\tau}_{2_0} = \frac{1}{w_{2_0}}\arccos\left(\frac{(p_2 w_{2_0}^2 - p_4 w_{2_0}^4 - p_0)(q_4 w_{2_0}^4 - q_2 w_{2_0}^2 + q_0) + (q_3 w_{2_0}^3 - q_1 w_{2_0})(p_3 w_{2_0}^3 - w_{2_0}^5 - p_1 w_{2_0})}{p\gamma\delta\left((q_4 w_{2_0}^4 - q_2 w_{2_0}^2 + q_0)^2 - (q_1 w_{2_0} - q_3 w_{2_0}^3)^2\right)}\right).$$

**Proof.**

Let $\lambda = iw, w > 0$ be a root of equation (22) to obtain

$$P(\lambda, \tau_2) = w^5 i + p_4 w^4 - p_3 w^3 i - p_2 w^2 + p_1 w i + p_0$$

$$+ (q_4 w^4 - q_3 w^3 i - q_2 w^2 + q_1 w i + q_0) p\gamma\delta e^{-iw\tau_2}.$$

Using Euler expansion and separating real and imaginary parts, we obtain

$$p\gamma\delta((q_4 w^4 - q_2 w^2 + q_0)\cos w\tau_2 + (q_1 w - q_3 w^3)\sin w\tau_2) = p_2 w^2 - p_4 w^4 - p_0,$$

$$p\gamma\delta((q_1 w - q_3 w^3)\cos w\tau_2 + (q_4 w^4 - q_2 w^2 + q_0)\sin w\tau_2) = p_3 w^3 - w^5 - p_1 w. \qquad (23)$$

Eliminating $\tau_2$ from equation (23), by squaring and adding these two equations and put $w^2 = z$, we obtain the Hopf frequency below:

$$z^5 + A_4 z^4 + A_3 z^3 + A_2 z^2 + A_1 z + A_0 = 0, \qquad (24)$$

with coefficients in (24) in Appendix A.3. Let's denote

$$g(z) = z^5 + A_4 z^4 + A_3 z^3 + A_2 z^2 + A_1 z + A_0 = 0.$$

Since $\lim_{z \to +\infty} g(z) = +\infty$ and $A_0 < 0$, then equation (24) has at least one positive root. Assuming equation (24) has $\tilde{n}$ positive roots, given by $\tilde{n}(1 \leq \tilde{n} \leq 5)$, denote by $z_1 < z_2 < \ldots z_{\tilde{n}}$, respectively. Then, equation (24) has $\tilde{n}$ positive roots if

$$w_1 = \sqrt{z_1}, w_2 = \sqrt{z_2}, \ldots, w_{\tilde{n}} = \sqrt{z_{\tilde{n}}}.$$

From (23), the corresponding $\tau_{2_n} > 0$, for which the characteristic equation (13) has a pair of purely imaginary roots is derived to have

$$\cos(w\tau_2) = \frac{(p_2 w^2 - p_4 w^4 - p_0)(q_4 w^4 - q_2 w^2 + q_0) + (q_3 w^3 - q_1 w)(p_3 w^3 - w^5 - p_1 w)}{(q_4 w^4 - q_2 w^2 + q_0)^2 + (q_3 w^3 - q_1 w)(q_1 w - q_3 w^3) p\gamma\delta}.$$

Thus, denoting

$$\tau_{2_n}^{(k)} = \frac{1}{w_n} \arccos\left(\frac{(p_2 w_n^2 - p_4 w_n^4 - p_0)(q_4 w_n^4 - q_2 w_n^2 + q_0) + (q_3 w_n^3 - q_1 w_n)(p_3 w_n^3 - w_n^5 - p_1 w_n)}{(q_4 w_n^4 - q_2 w_n^2 + q_0)^2 + (q_3 w_n^3 - q_1 w_n)(q_1 w_n - q_3 w_n^3) p\gamma\delta}\right) + \frac{2\pi(k-1)}{w_n}, \quad (25)$$

where

$n = 1, 2, \ldots \tilde{n}, k = 1, 2, \ldots$, then $\pm i w_n$ are a pair of purely imaginary roots of the equation (13). This allows us to define the Hopf–bifurcation threshold time delay value as

$$\tau_{2_0} = \frac{1}{w_{2_0}} \arccos\left(\frac{(p_2 w_{2_0}^2 - p_4 w_{2_0}^4 - p_0)(q_4 w_{2_0}^4 - q_2 w_{2_0}^2 + q_0) + (q_3 w_{2_0}^3 - q_1 w_{2_0})(p_3 w_{2_0}^3 - w_{2_0}^5 - p_1 w_{2_0})}{p\gamma\delta\left((q_4 w_{2_0}^4 - q_2 w_{2_0}^2 + q_0)^2 - (q_1 w_{2_0} - q_3 w_{2_0}^3)^2\right)}\right). \quad (26)$$

This completes the proof.

**Proposition 4.3** *If conditions*
$5w_{2_0}^4(q_1 + 2q_2 w_{2_0}) + (3p_3 w_{2_0}^2 + p_1)(3q_3 w_{2_0}^2 + 4q_3 w_{2_0}^3) > 5w_{2_0}^4(3q_3 w_{2_0}^2 + 4q_4 w_{2_0}^3) + (q_1 + 2q_2 w_{2_0})(3p_3 w_{2_0}^2 + p_1),$

$\dfrac{q_3 w_{2_0}^2}{q_1} > 1, \dfrac{w_{2_0}^4 q_4 + q_0}{q_2 w_{2_0}^2} > 1$ *holds, such that* $\Phi_1(w_{2_0}) > 0$, *then system* (1) *undergoes a Hopf–bifurcation at the endemic equilibrium point as* $\tau_2$ *increases through* $\tau_{2_0}$, *where expressions of* $\Phi_1(w_{2_0})$ *satisfies equation* (29).

**Proof (Transversality condition for Hopf–bifurcation)**

In order to establish whether the endemic equilibrium point $P^*$ actually under goes a Hopf–bifurcation at $\tau_2 = \tau_{2_0}$, we let $\lambda(\tau_2) = \beta(\tau_2) + iw(\tau_2)$ be a root of equation (13) near $\tau_2 = \tau_{2_0}^{(k)}$ and $\beta(\tau_2)^{(k)} = 0$, as

$w(\tau_2)^{(k)} = w_{2_0}$. Making a substitution into the L.H.S of equation (13) and taking a derivative with respect to $\lambda$, we have

$$\frac{d\tau_2}{d\lambda} = \frac{(5\lambda^4 + 4p_4\lambda^3 + 3p_3\lambda^2 + 2p_2\lambda + p_1)e^{\mu\lambda\tau_2}}{(q_4\lambda^5 + q_3\lambda^4 + q_2\lambda^3 + q_1\lambda^2 + q_0\lambda)p\gamma\delta} + \frac{(4q_4\lambda^3 + 3q_3\lambda^2 + 2q_2\lambda + q_1)}{p\gamma\delta(q_4\lambda^5 + q_3\lambda^4 + q_2\lambda^3 + q_1\lambda^2 + q_0\lambda)} - \frac{\tau_2}{\lambda}. \quad (27)$$

Computing the Sign of $\dfrac{d[Re(\lambda)]}{d\tau_2}$, by differentiating the characteristic equation (13) with respect to $\tau_2$ and evaluating (27) at $\tau_2 = \tau_{2_0}$ with $\lambda = iw_{2_0}$ and expressing $\sin(w_{2_0}\tau_{2_0})$ and $\cos(w_{2_0}\tau_{2_0})$, we obtain sign

$$\text{sign}\left[\frac{d(Re\lambda)}{d\tau_2}\right]_{\tau_2=\tau_{2_0}} = \text{sign}\left[Re\left(\frac{d\lambda}{d\tau_2}\right)^{-1}\right]_{\lambda=iw_{2_0}},$$

$$= \text{sign}\left[Re\frac{f_1\cos d_0 + f_2\sin d_0}{g_1 + ig_2} + Re\frac{i(f_3\cos d_0 + f_4\sin d_0)}{(g_1 + ig_2)} + Re\frac{f_5}{g_1 + ig_2} - Re\frac{\tau_2}{iw_{2_0}}\right],$$

$$= \text{sign}\left[Re\frac{g_1(f_1\cos d_0 + f_2\sin d_0) - ig_2(f_1\cos d_0 + f_2\sin d_0)}{g_1^2 + g_2^2}\right] +$$

$$\text{sign}\left[Re\frac{g_2(f_2\cos d_0 + f_4\sin d_0) + ig_1(f_3\cos d_0 + f_4\sin d_0)}{g_1^2 + g_2^2}\right]$$

$$+ \text{sign}\left[Re\frac{f_5 g_1}{g_1^2 + g_2^2}\right], \quad (28)$$

with coefficients in Appendix A.4

By **Remark 4.1**, equation (28), gives

$$\text{sign}\left[\frac{g_1(\sqrt{f_1^2 + f_2^2}(\sin(d_0 + \Psi_0))) + g_2\sqrt{f_2^2 + f_4^2}(\sin(d_0 + \Psi_1)) + f_5 g_1}{g_1^2 + g_2^2}\right],$$

With $\Psi_0 = \arctan\dfrac{f_1}{f_2}, \Psi_1 = \arctan\dfrac{f_2}{f_4}$ and $f_2 \neq 0, f_4 \neq 0$

Let

$$\Phi_1(w_{2_0}) = g_1\sqrt{f_1^2 + f_2^2}\left(\sin(d_0 + \Psi_0)\right) + g_2\sqrt{f_2^2 + f_4^2}\left(\sin(d_0 + \Psi_1)\right) + f_5 g_1. \quad (29)$$

If $\Phi_1(w_{2_0}) > 0$, with $(d_0 + \Psi_{\{i=0,1\}}) \in (\pi, \frac{\pi}{2}]$, then $\text{sign}[\frac{d(\text{Re}\lambda)}{d\tau_2}]_{\tau_2 = \tau_{2_0}} > 0$, hence the transversality condition holds and the system undergoes Hopf–bifurcation.

### 4.3 Delay in latent period and seeking medical care ($\tau_1 = \tau_2 = \tau > 0$)

Making a substitution of $\tau_1 = \tau_2 = \tau$ in equation (13), we get

$$g(\lambda, e^{-\lambda\tau}) = \lambda^5 + k_4\lambda^4 + k_3\lambda^3 + k_2\lambda^2 + k_1\lambda + k_0 + ((s_4)\lambda^4 + s_3\lambda^3 + s_2\lambda^2 + s_1\lambda + s_0)e^{-\lambda\tau}$$
$$+(s_3'\lambda^3 + s_2'\lambda^2 + s_1'\lambda + s_0')e^{-2\lambda\tau} = 0. \tag{30}$$

with

$$s_4 = (\gamma + \delta)e^{-\mu\tau}, s_3 = (\gamma l_3 + m_3\delta)e^{-\mu\tau}, s_2 = (l_2\gamma + m_2\delta)e^{-\mu\tau}, s_1 = (l_1\gamma + m_1\delta)e^{-\mu\tau},$$
$$s_0 = (l_0\gamma + m_0\delta)e^{-\mu\tau}, s_3' = (\gamma\delta)e^{-2\mu\tau}, s_2' = n_2\delta\gamma e^{-2\mu\tau}, s_1' = n_1\gamma\delta e^{-2\mu\tau}, s_0' = n_0\gamma e^{-2\mu\tau}$$

In order to examine whether or not the endemic equilibrium loses stability and undergoes Hopf–bifurcation as an outcome with inclusion of the time delays, a pair of purely imaginary root of the transcendental equation (30) is found. Suppose the pair of the imaginary root is given as $\lambda = iv$ with infection rate oscillation frequency ($v > 0$), using Euler's expansion and making a substitution into equation (30), separating real and imaginary parts, we obtain:

$$g_0 \cos v\tau + g_1 \sin v\tau + g_2 \sin 2v\tau = G_1, \tag{31}$$

$$-g_1 \cos v\tau + g_0 \sin v\tau + g_3 \sin 2v\tau = G_2. \tag{32}$$

where
$g_0 = s_1 v - s_3 v^3, g_1 = s_2 v^2 - s_4 v^4 - s_0, g_2 = s_2 v^2, g_3 = s_3' v^3 + s_1' v, G_1 = v^5 + (k_3 + s_3 + s_3')v^3 - (k_1 + s_1')v,$
$G_2 = (k_2 + s_2')v^2 - (k_4 v^4 + k_0 + s_0').$

Squaring and adding equation (31) and (32), we get following equation:

$$g_0^2 + g_1^2 - G_1^2 - G_2^2 = -\frac{1}{2}(g_2^2 + g_3^2)(1 - \cos 4v\tau). \tag{33}$$

suppose $|\cos 4v\tau| < 1$, , equation (33) leads to

$$G_1^2 + G_2^2 - (g_0^2 + g_1^2) = 0, \tag{34}$$

which reduces to

$$v^{10} + (2(k_3 + s_3 + s_3') + k_4^2 - s_4^2)v^8 + ((k_3 + s_3 + s_3')^2 + 2(k_1 + s_1') - 2k_4(k_2 + s_2') + 2s_2s_4 - s_3^2)v^6$$

$$+ (2(k_1 + s_1')(k_3 + s_3 + s_3') + 2k_4(k_0 + s_0') + 2s_1s_3 - (s_2^2 + 2s_0s_4))v^4$$

$$+ ((k_1 + s_1')^2 + 2s_0s_2 - 2(k_2 + s_2')(k_0 + s_0') - s_1^2)v^2 + (k_0 + s_0')^2 = 0. \quad (35)$$

Let $z = v^2$ such that we obtain equation (35) in terms of $z$:

$$L(z) = z^5 + u_4 z^4 + u_3 z^3 + u_2 z^2 + u_1 z + u_0, \quad (36)$$

with

$$u_4 = 2(k_3 + s_3 + s_3') + k_4^2 - s_4^2, u_3 = (k_3 + s_3 + s_3')^2 + 2(k_1 + s_1^1) - 2k_4(k_2 + s_2^1) + (2s_2s_4 - s_3^2),$$
$$u_2 = 2(k_1 + s_1')(k_3 + s_3 + s_3') + 2k_4(k_0 + s_0') + 2s_1s_3 - (s_2^2 + 2s_0s_4), u_0 = (k_0 + s_0')^2,$$
$$u_1 = (k_1 + s_1')^2 + 2s_0s_2 - 2(k_2 + s_2')(k_0 + s_0') - s_1^2.$$

Since equation (36), has a high degree polynomial we compute the eigenvalues numerically by using parameter values in Table 2. The resulting polynomial is

$$z^5 - 295.18z^4 - 130.18z^3 + 92.52z^2 - 0.15038z + 2.6 \times 10^{-12} = 0.$$

Therefore, the following eigenvalues are obtained:

$z_1 = 295.62, z_2 = 0, z_3 = 0.001629, z_4 = 0.38024, z_5 = -0.8212$

We observe that there is only one negative real root which does not guarantee stability of model (1) in the presence of time delays $\tau = \tau_1 = \tau_2 > 0$ thus by Lemma 4.2 there exists a pure imaginary root $w_c$ such that a critical time delay $\tau_c$ is achieved for which there is death or birth of period oscillations (Hopf–bifurcation). Equation (33) yields

$$\tau_c = (\frac{1}{4v_0} \arccos \frac{g_2^2 + g_3^2 + 2(g_0^2 + g_1^2 - (G_1^2 + G_2^2))}{g_2^2 + g_3^2}) + \frac{j\pi}{2v_0}; j = 0,1,3,\ldots \quad (37)$$

with $\lambda = iv$ (a purely imaginary root of equation (30)), if condition $g_0^2 + g_1^2 = G_1^2 + G_2^2$ and $\tau \in [0, \tau_c)$ holds.

Without loss of generality, let $v_0$ represent the value $v_0$ corresponding to $\tau_c$. We thus state the result below:

**Proposition 4.4.** *If condition $g_0^2 + g_1^2 = G_1^2 + G_2^2$ holds, then the chronic steady state $P^*$ is locally asymptotically stable for $\tau \in [0, \tau_c)$, unstable when $\tau > \tau_c$ and undergoes a Hopf–bifurcation.*

## 5 Numerical simulation and results

In this Section, we use MATLAB dde23 function to obtain numerical simulations and graphical representations of model (1) to supplement the analytical solutions in Section 4. Parameter values in Table 2 are used in the simulation.

The positive endemic equilibrium is $P^* = (S^*, V^*, E^*, C^*, I^*) = (2099, 6, 54, 2, 100)$. In the absence of delays $\tau_1 = \tau_2 = 0$, the characteristic polynomial equation (14) is

$$\lambda^5 + 0.7364\lambda^4 - 148.4007\lambda^3 - 4.9408\lambda^2 - 0.3965\lambda - 0.0001806.$$

The corresponding eigenvalues are;

$$\lambda_1 = 11.8366, \lambda_2 = -0.000472, \lambda_3 = -12.5357, \lambda_{4,5} = -0.01641 \pm 04885i.$$

Therefore, since the eigenvalues have one positive root and four negative roots, the endemic equilibrium changes state of stability from unstable to stable thus under goes a Hopf–bifurcation (see Figure 3). This implies that as time approaches infinity, the partial populations are stable and pneumococcal pneumonia can no longer cause harm individual

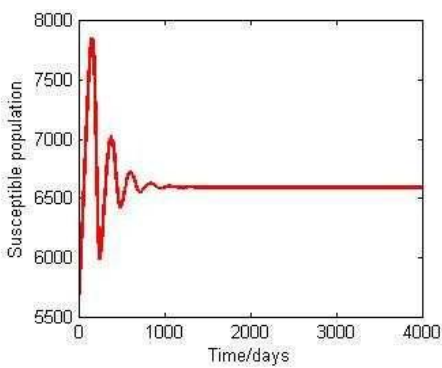 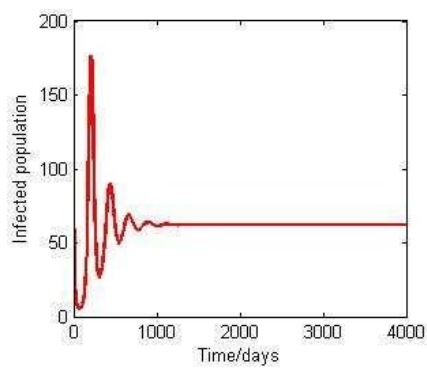

(a)          (b)

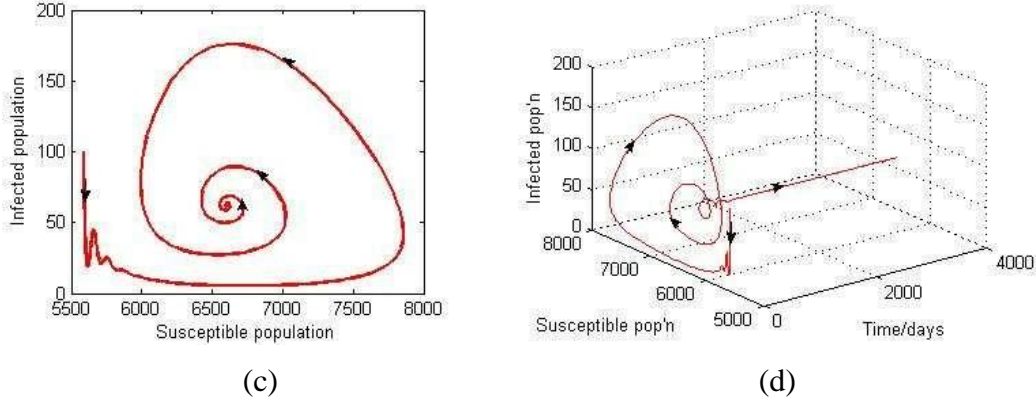

(c)  (d)

Figure 3: (a& b) Stability of the endemic equilibrium showing Hopf bifurcation, with initial variables: $S(0) = 5586$, $V(0) = 22$, $C(0) = 64$, $I(0) = 11$, $E(0) = 100$. (c& d) The evolution of the infected, the susceptible and corresponding I-S portrait and 3-D phase trajectories,
(with $R_0 = 15.4, R_0^u = 15.14, R_0^v = 0.271$ parameters:
$\mu = 2.0547 \times 10^{-4}, \phi = 3.574 \times 10^{-2}$. The rest of the parameters remain fixed as in Table 2

The numerical simulation of equation (15) yields the characteristic roots as;

$$\lambda_1 = 0, \lambda_2 = 14.4621i, \lambda_3 = -14.4416, \lambda_4 = \pm 0.00041 + 0.3771i, \lambda_5 = \pm 0.0579 + 0.1335i.$$

As $\tau_1$ increase from zero, there is a value $\tau_{I_0} > 0$ such that the endemic equilibrium is stable for $\tau_1 = [0, \tau_{I_0}]$ and unstable for $\tau_1 > \tau_{I_0}$. At this critical value, the endemic equilibrium loses stability and Hopf bifurcation arises. The real positive root is $w_{I_0} = 14.4621$ and the critical time delay $\tau_{I_0} = 0.109$ of a day $\approx$ 3 hrs.

Figure 4 shows the evolution of the susceptible and infected population of system (1). The low and high peaks in the number of susceptible and infected individuals indicate the season peak of the disease. If $\tau_1 < \tau_{10} = 0.109$ of a day $\approx$ 3 hrs, the partial populations of the susceptible and the infected are stable whereas if $\tau_1 > \tau_{10} = 0.109$ of a day $\approx$ 3hrs, the populations are unstable and it's hard to predict the future pattern of the disease prevalence.

The numerical computation of equation (22) yields eigenvalues

$$\lambda_1 = 0, \lambda_2 = 0.06508i, \lambda_3 = -0.06522, \lambda_4 = -12.038, \lambda = -12.3263.$$ The positive root

$w_{20} = 0.06508$ and the critical time delay $\tau_{2_0} = 26$, hence system (1) is stable for $\tau_2 < 26$ days and unstable for $\tau_2 > \tau_{2_0}$. A characteristic polynomial (30) corresponding to two delays is solved to give the eigenvalues as; $\lambda_1 = 0, \lambda_2 = -17.1963, \lambda_3 = -0.6166, \lambda_4 = -0.04036, \lambda_5 = 0.9026i$, the real positive root $w_c = 0.9062$ and the critical time delay $\tau_c = 2.069$ days.

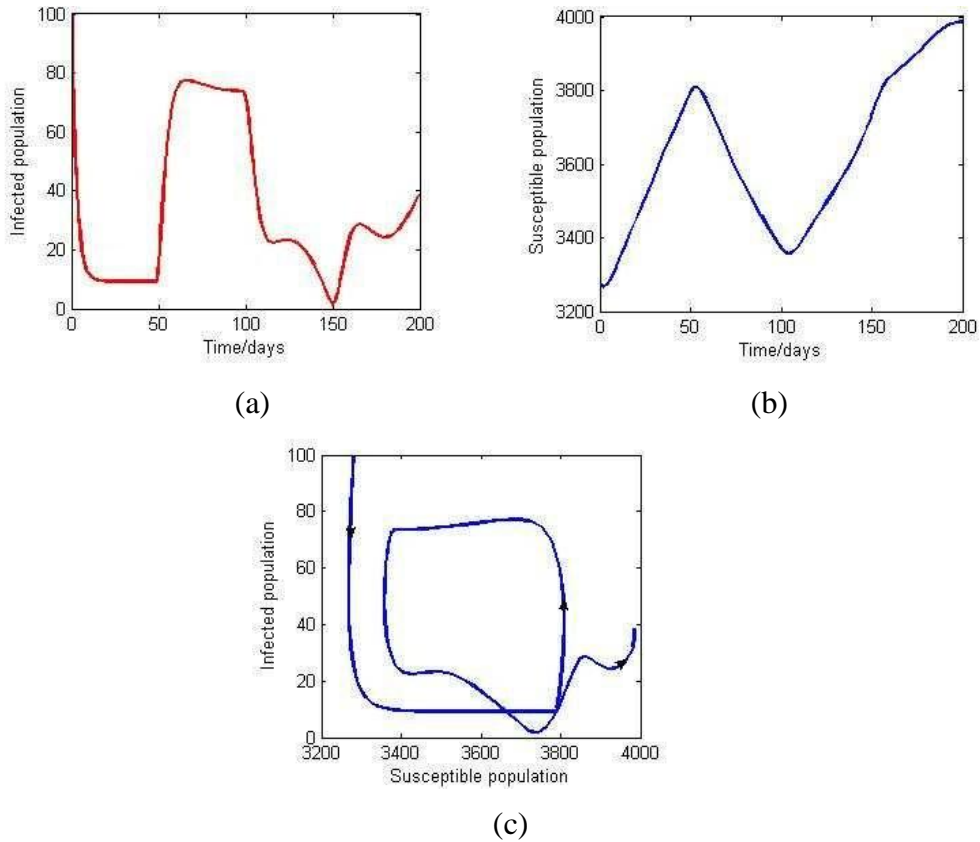

(a)

(b)

(c)

Figure 4: Simulation of model (1) for $\tau_2 = 0$ and $\tau_1 > 0$, with initial variable values: (S (0), V (0), E (0), C (0), I (0)) = (3280, 30, 10, 10, 100). The rest of the parameters are as in Table 2.

Figure 5 depicts the time series solution approaching their equilibrium point as time approaches infinity. This confirms the stability of the system when the value of time delay is less than $\tau_c = 2.069$ days and instability of the system if $\tau > \tau_c = 2.069$ days (see Proposition 4.3).

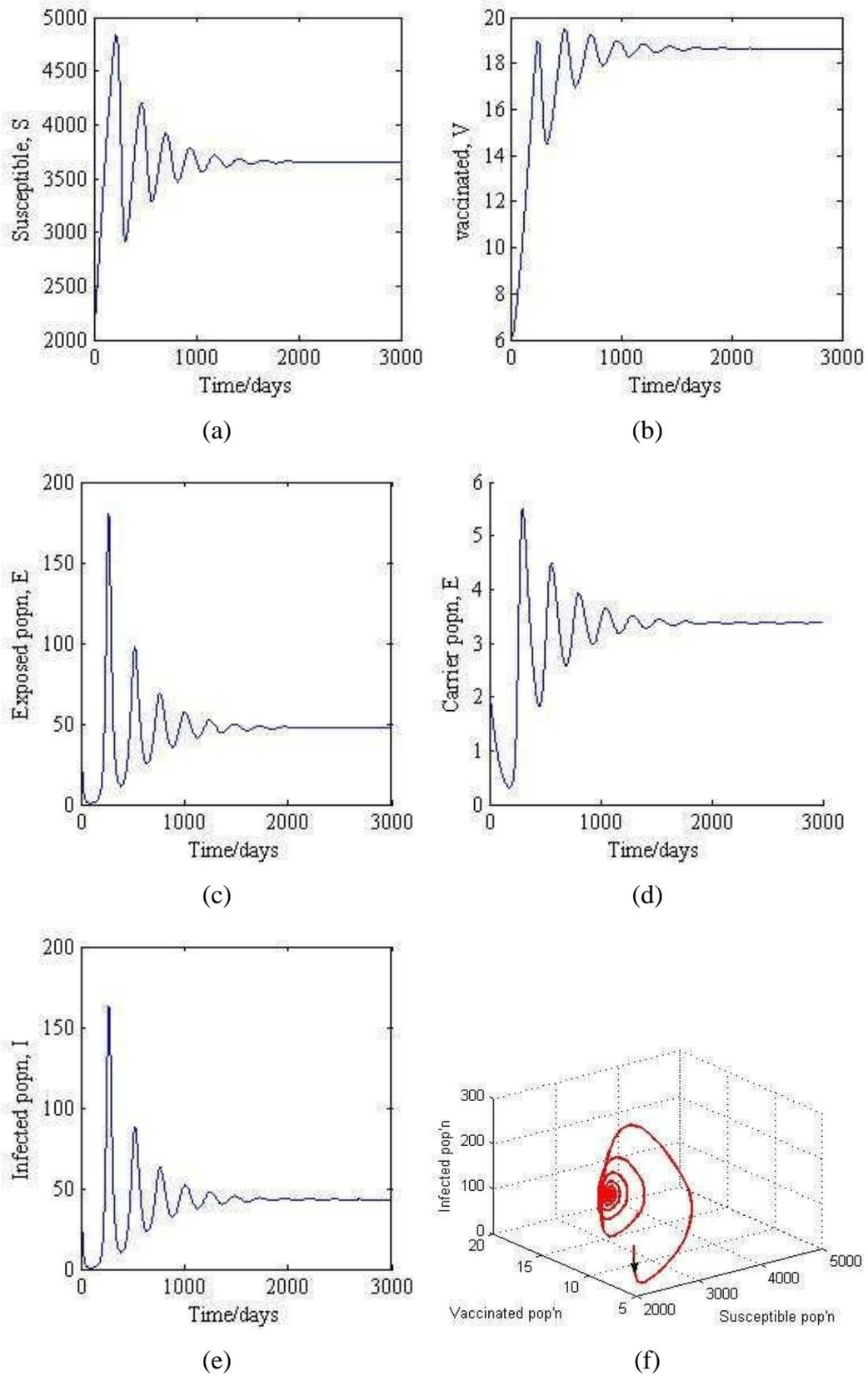

Figure 5: Stability of the endemic equilibrium $P^*$ for $\tau_1 = \tau_2 = 2$ days. The rest of the parameters are as in Table 2.

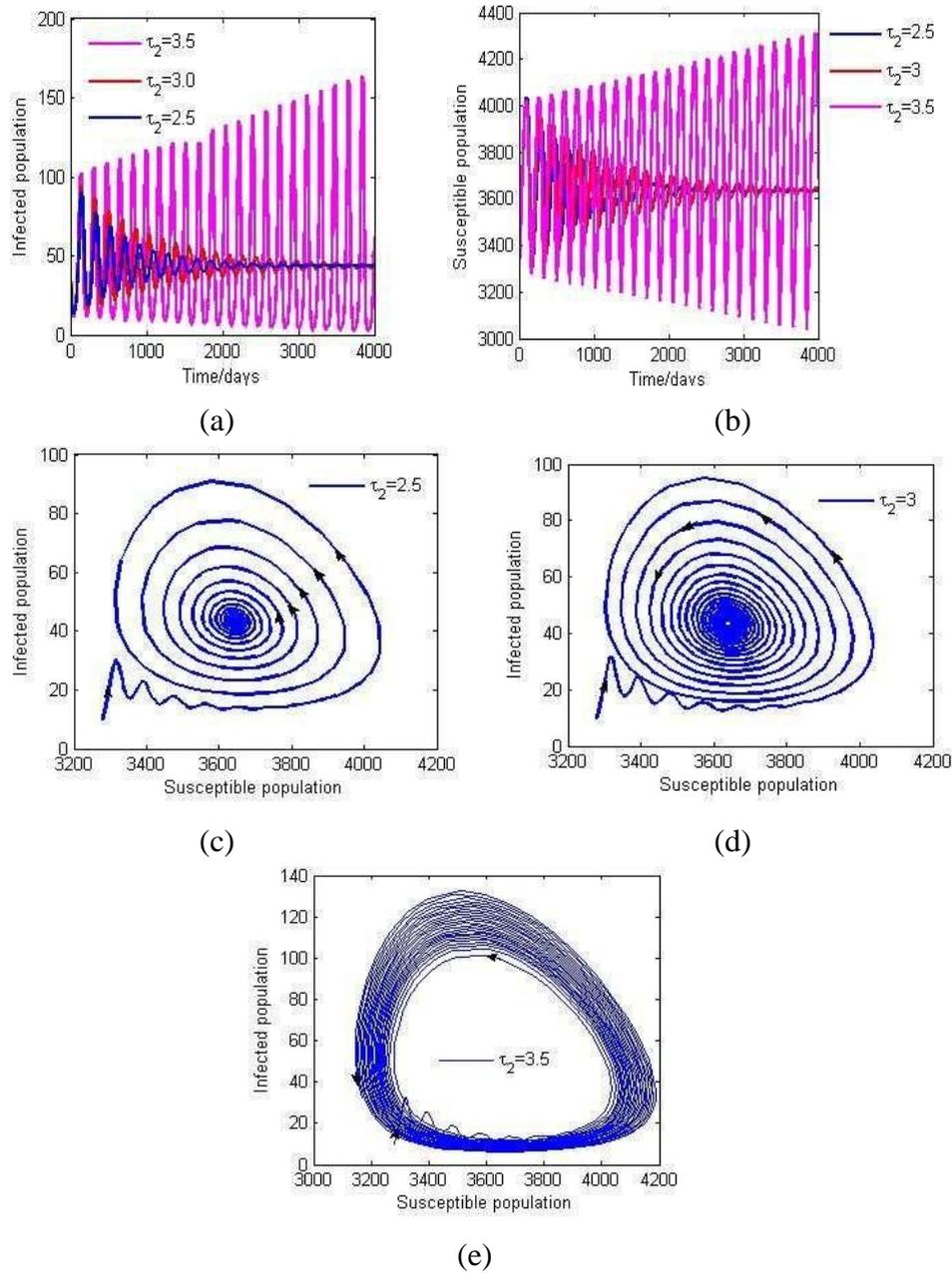

Figure 6: The effect of varying $\tau_2$ on the dynamics of model (1). The delay $\tau_2$ was chosen as $\tau_2 = 2.5, 3, 3.5$. All other parameters remain as stated in Table 2

To explore the effect of time delay $\tau_2$ on pneumococcal pneumonia, we fix time delay $\tau_1 = 3$ days, and the parameter $\tau_2$ is varied (Figure 6). The rate of convergence to stability of the endemic equilibrium point is attained with a reduction in the delay and a divergence is due to an increase in the delay that results into instability of the system. This gives rise to Hopf–bifurcation phenomenon.

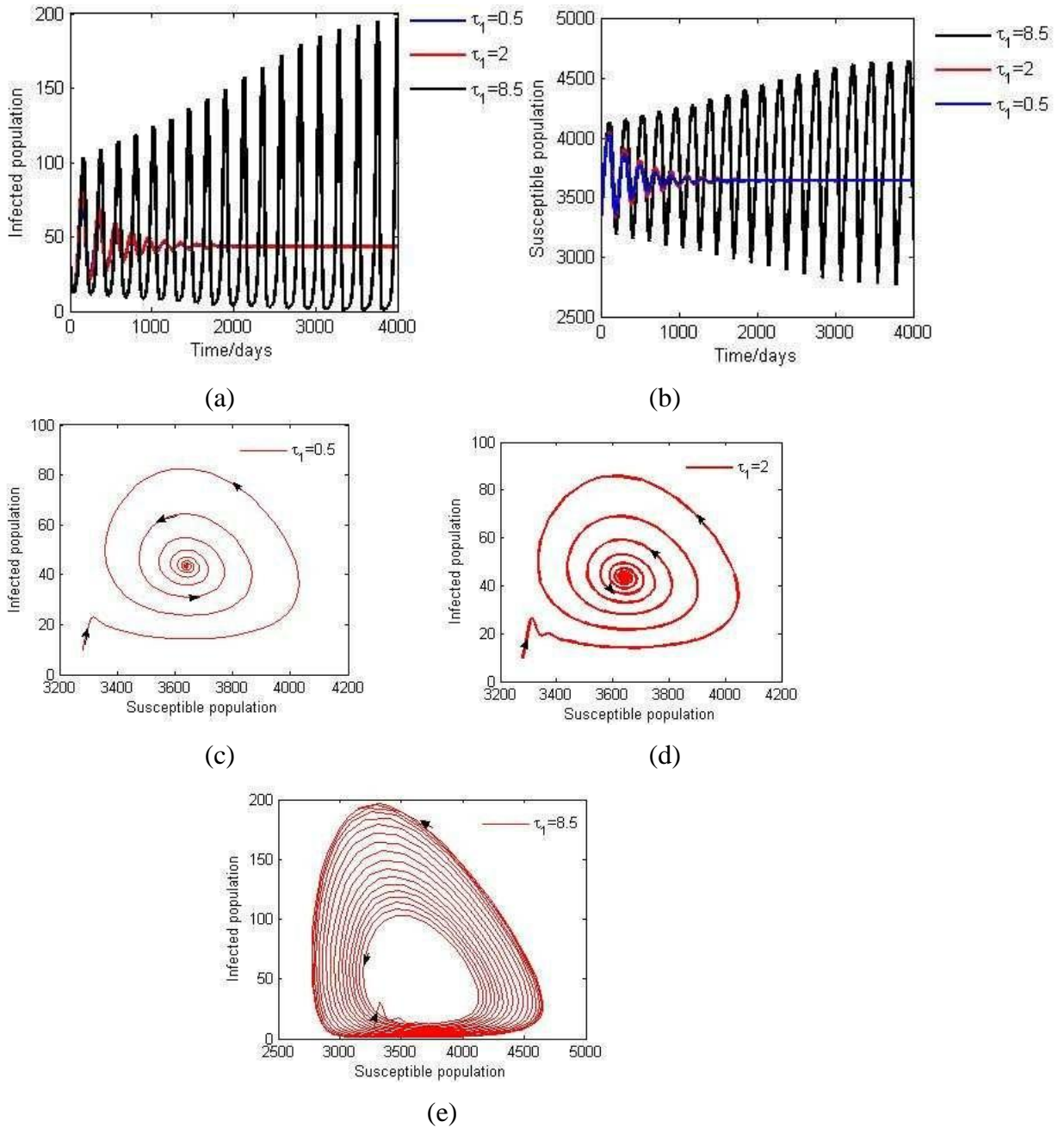

Figure 7: The effect of varying time delay $\tau_1$ on the dynamics of model (1). The delay $\tau_1$ was chosen as $\tau_1 = 0.5, 2, 8.5.$ All the parameter values are the same as in Table 2

In Figure 7, time delay $\tau_2$ is fixed at 2 days in order to study the effect of time delay $\tau_1$ on model (1). We observe an increase in the magnitude of the amplitude of oscillations as $\tau_1$ increases, thus divergence from the endemic equilibrium occurs leading to unstable state. This implies that the disease will persist in the population with increased delays if there is no intervention instituted to reduce the delays. On the other hand, a decrease in $\tau_1$ guarantees the asymptotic stability of the endemic equilibrium which implies the disease can be eradicated from the population.

## 6 Discussion

In this paper, we propose and analyze a mathematical model of pneumococcal pneumonia with time delays. We derive the control reproductive ratio $R_0$. The results show that, without delays ($\tau_1 = \tau_2 = 0$), the disease–free equilibrium $P_0$ is locally asymptotically stable if the control reproductive ratio $R_0 < 1$, whenever conditions $(\mu + \zeta)(\mu + \nu) > \zeta\nu$ and $R_0^\nu < 1$ hold, and unstable if $R_0 > 1$.

The analysis of model (1) is done. The results show that the endemic equilibrium is locally stable without delays and stable if the delays are under conditions. The transversality conditions for the existence of Hopf–bifurcation are stated and proved for three cases; (1) $\tau_1 = 0, \tau_2 = \tau > 0$, (2) $\tau_1 = \tau > 0, \tau_2 = 0$ and (3) $\tau_1 = \tau_2 = \tau > 0$. Critical values at which Hopf–bifurcation occur have been obtained. The results show that at critical values $\tau_{1_0} = 0.109 \approx 3$ hrs, $\tau_{2_0} = 26$ days and $\tau_c = 2.069$ days, the endemic equilibrium losses stability.

We investigated the effect of two delays $\tau_1$ and $\tau_2$ on the stability of model (1). Basing on the numerical simulations obtained in this paper, we found out that when $\tau_1$, $\tau_2$ are below the critical values $\tau_{1_0}$ and $\tau_{20}$ respectively, model (1) is asymptotically stable. Which implies that the number of individuals in the five subpopulations will be in ideal equilibrium and prevalence of pneumococcal pneumonia can easily be controlled. Conversely, if the value of the delays $\tau_1$, $\tau_2$ are greater than the critical values $\tau_{1_0}$ and $\tau_{20}$ respectively, a Hopf–bifurcation arises this phenomenon suggests persistent of pneumococcal pneumonia in the population. The number of individuals in the five subpopulations of model (1) will fluctuate periodically, this is not helpful, effort should be put to control such a phenomenon.

Longer time delays destabilize the system and give rise to Hopf bifurcations. This explains the oscillatory seasonal change of pneumococcal pneumonia disease in human population whose immune systems are weak. Therefore, measures to reduce delays in latent and seeking medical care during pneumococcal pneumonia epidemic should be prioritized. The results herein could be helpful to direct future research of bacterial infections that become severe in individuals that have history of exposure to viral infections such as influenza A virus.

**Data availability**

Data supporting this time delay model are from previously published research articles, parameter values have been cited in Table 2.

**Acknowledgments**

The authors are grateful to Prof. Roberto Barrio (Editor) and the anonymous referees for their significant comments and suggestions, which have helped to improve the manuscript. We would like to thank Pan African University Institute of Basic Sciences Technology and Innovation (PAUSTI), Busitema University and Uganda Martyrs' University for the financial support.

**Conflict of interests**

The authors declare that there is no conflict of interests concerning the publication of this paper.

A Detailed mathematical coefficient terms in the paper with corresponding computed values obtained by using parameters from Table 2

**Appendix**

**A.1 Coefficient terms in the transcendental equation (13)**

$k_4 = -(a_1 + a_5 + a_8 + a_{11} + a_{14})$,

$k_3 = (a_8(a_1 + a_5) + a_{11}(a_5 + a_1 + a_8 + a_{14}) + a_{14}(a_5 + a_1 + a_8) + a_1 a_5 - a_{12} a_{13} - a_2 a_4)$,

$k_2 = (a_2 a_4 (a_8 + a_{11} + a_{14}) + a_{12} a_{13}(a1 + a_5 + a_8) - a_1 a_5(a_8 + a_{11} + a_{14}) - a_1 a_8(a_{11} + a_{14}) - a_{11}(a_5 a_8 + a_1 a_{14}) - a_{14}(a_5 a_8 + a_5 a_{11}) + a_8 a_{11}))$,

$k_1 = (a_8 a_{11}(a_1 a_5 - a_2 a_4) + a_1 a_5(a_8 a_{14} + a_{11} a_{14}) - a_2 a_4 a_{14}(a_8 + a_{11}) + a_{12} a_{13}(a_2 a_4 - a_1 a_5 - a_1 a_8 - a_5 a_8) + a_{11} a_{14}(a_1 a_8 + a_5 a_8)), k_0 = a_8(a_{11} a_{14} - a_{12} a_{13})(a_2 a_4 - a_1 a_5)$,

$l_3 = -(a_5 + a_{11} + a_{14} + a_1 + a_9)$,

$l_2 = (a_1(a_5 + a_{11} + a_{14}) + a_5(a_{11} + a_{14})) + (a_1 a_9 - a_3 a_7 + a_5 a_9 + a_9 a_{11})$,

$l_1 = (a_2 a_4(a_{14} + a_{11}) - a_1 a_5(a_{11} - a_{14}) - a_{11} a_{14}(a_1 - a_5) + a_{12} a_{13}(a_5 + a_1)) + (a_3 a_7(a_5 + a_{11}) + a_2(a_4 a_9 - a_6 a_7) - a_9 a_{11}(a_1 + a_5) - a_1 a_5 a_9)$,

$l_0 = (a_{12} a_{13}(a_2 a_4 - a_1 a_5) - a_{11} a_{14}(a_2 a_4 + a_1 a_5))) + a_7 a_{11}(a_2 a_6 - a_3 a_5) + a_9 a_{11}(a_1 a_5 - a_2 a_4)), m_3 = (a_1 + a_5 + a_8 + a_{11})$,

$m_2 = (a_1 a_5 - a_2 a_4 + a_1 a_8 + a_1 a_{11} + a_5 a_8 + a_5 a_{11} + a_8 a_{11})$,

$m_1 = (a_2 a_4 a_8 - a_1 a_5 a_{11} + a_2 a_4 a_{11} - a_1 a_8 a_{11} - a_5 a_8 a_{11} - a_1 a_5 a_8))$,

$m_0 = a_8 a_{11}(a_1 a_5 - a_2 a_4), n_2 = ((a_5 + a_{11} + a_1) + a_{11} a_{14} - a_{12} a_{13} - a_2 a_4), n_1 = (a_{11}(a1 + a_5) + a_1 a_5 - a_2 a_4)$,

$n_0 = a_{11}(a_2 a_4 - a_1 a_5))$.

Hence $a_1 = -0.01218, a_2 = 5.479 \times 10^{-4}, a_3 = -0.1764, a_4 = 2.53 \times 10^{-5}, a_5 = 0.008057,$
$a_6 = -0.0003596, a_7 = 0.0101, a_9 = 445, a_{10} = 0.005455, a_{11} = -0.01302,$

$a_{12} = 0.0003596, a_{13} = 0.01096, a_{14} = -0.03777, a_{15} = -0.3319, a_{16} = 0.33196, a_{17} = 0.3279, k_4 = -0.07308, k_3 = 0.001759,$

$k_2 = -0.00008172, k_1 = 0.0005897, k_0 = 9.833 \times 10^{-11}, l_3 = -445.2, l_2 = -14.81, -1.1915, l_0 = -0.0005686,$

$m_3 = -0.03531, m_2 = 0.0004299, m_1 = 0.00006202, m_0 = 00002674, n_2 = -0.3277, n_1 = 0.0003616, n_0 = 0.000001277.$

## A.2 Coefficient terms in the characteristic equation (14)

$b_4 = k_4 + \gamma + \delta, b_3 = k_3 + l_3\gamma + m_3\delta, b_2 = k_2 + l_2\gamma + m_2\delta + n_2\gamma\delta, b_1 = k_1 + l_1\gamma + m_1\delta + n_1\gamma\delta,$
$b_0 = l_o\gamma + m_0\delta + n_0\gamma\delta,$

hence $b_4 = 0.7364, b_3 = -148.4007, b_2 = -4.9408, b_1 = -0.3965, b_0 = -0.0001806.$

## A.3 Coefficient terms in equation (24)

$A_4 = (p_2^2 - p^2\gamma^2\delta^2 q_4 - 2p_3), A_3 = (2p_1 + p_3^2 - 2p_2 p_4 - p^2\gamma\delta^2(q_3^2 - 2q_4 q_2)),$

$A_2 = (p_2^2 - 2p_1 p_2 - p^2\gamma^2\delta^2(2q_4 q_0 + q_2^2 - 2q_1 q_3)), A_1 = p_1 - p^2\gamma^2\delta^2(q_1 - 2q_2 q_0), A_0 = -q_0^2 p^2\gamma^2\delta^2,$

hence $A_4 = 296.9, A_3 = 22018, A_2 = 0.3754, A_1 = -0.3966, A_0 = -2585 \times 10^{-11}.$

## A.4 Coefficient terms in Transversality condition of equation (28)

$d_0 = w_{2_0}\tau_2, f_1 = 5w_{2_0}^4 - (3p_3 w_{2_0}^2 + p_1), f_2 = 4p_4 w_{2_0}^3 - 2p_2 w_{2_0}, f_3 = 2p_2 w_{2_0} - 4p_4 w_{2_0}^3,$

$f_4 = 5w_{2_0}^4 - 3p_3 w_{2_0}^2, g_1 = p\gamma\delta(q_3 w_{2_0}^4 - q_1 w_{2_0}^2), g_2 = p\gamma\delta(q_4 w_{2_0}^5 + q_0 w_{2_0} - q_2 w_{2_0}^3),$

$f_5 = q_1 + 2q_2 w_{2_0} - (3q_3 w_{2_0}^2 + 4q_4 w_{2_0}^3),$

hence $p = 0.9939, p_4 = 0.4064, p_3 = -148.4, p_2 = 0.3333, p_1 = -0.3966, p_0 = -0.0001895,$

$q_4 = 0.3279, q_3 = 0.6280, q_2 = -0.003463, q_1 = 0.00004153, q_0 = 0.00002672.$

## B Computation of critical values

### B.1 Critical value for seeking medical care $\tau_{2_0}$

By applying L'Hopitals rule to the arccos function of equation (26), let

$$y = \arccos(Z)$$

$$Z = \frac{(p_2 w_{2_0}^2 - p_4 w_{2_0}^4 - p_0)(q_4 w_{2_0}^4 - q_2 w_{2_0}^2 + q_0) + (q_3 w_{2_0}^3 - q_1 w_{2_0})(p_3 w_{2_0}^3 - w_{2_0}^5 - p_1 w_{2_0})}{p\gamma\delta\left((q_4 w_{2_0}^4 - q_2 w_{2_0}^2 + q_0)^2 - (q_1 w_{2_0} - q_3 w_{2_0}^3)^2\right)},$$

$$\frac{d(y)}{dw_{2_0}} = -\frac{(2p_2 w_{2_0} - 4p_4 w_{2_0}^3 - p_0)(q_4 w_{2_0}^4 - q_2 w_{2_0}^2 + q_0) + (p_2 w_{2_0}^2 - p_4 w_{2_0}^4 - p_0)(4q_4 w_{2_0}^3 - 2q_2 w_{2_0})}{p\gamma\delta\left(2(q_4 w^4 - q_2 w^2 + q_0))(4q_4 w^3 - 2q_2 w) - 2(q_1 w - q_3 w^3)(q_1 - 3q_3 w^2)\sqrt{1-Z^2}\right)}$$

$$-\frac{(p_3 w_{2_0}^3 - w_{2_0}^5 - p_1 w_{2_0})(3q_3 w_{2_0}^2 - q_1) + (q_3 w_{2_0}^3 - q_1 w)(3p_3 w_{2_0}^2 - 5w_{2_0}^4 - p_1)}{p\gamma\delta\left(2(q_4 w_{2_0}^4 - q_2 w_{2_0}^2 + q_0))(4q_4 w_{2_0}^3 - 2q_2 w_{2_0}) - 2(q_1 w_{2_0} - q_3 w_{2_0}^3)(q_1 - 3q_3 w_{2_0}^2)\sqrt{1-Z^2}\right)} = 0.174$$